\documentclass[12pt]{article}

\usepackage[letterpaper,hmargin=1in,vmargin=1in]{geometry}

\usepackage{graphicx,amsmath,amsfonts}

\def\be{\begin{equation}}
\def\ee{\end{equation}}
\def\ba{\begin{eqnarray}}
\def\ea{\end{eqnarray}}
\def\ge{\mathrel{\raise.3ex\hbox{$>$\kern-.75em\lower1ex\hbox{$\sim$}}}}
\def\la{\mathrel{\raise.3ex\hbox{$<$\kern-.75em\lower1ex\hbox{$\sim$}}}}

\def\theequation{\thesection.\arabic{equation}}
\def\simgt{\mathrel{\raise.3ex\hbox{$>$\kern-.75em\lower1ex\hbox{$\sim$}}}}
\def\simlt{\mathrel{\raise.3ex\hbox{$<$\kern-.75em\lower1ex\hbox{$\sim$}}}}

\newcommand{\bi}[1]{\bibitem{#1}}

\newcommand{\nc}{\newcommand}

\nc{\gone}{\bar g_{\pi NN}^{(1)}}
\nc{\gzero}{\bar g_{\pi NN}^{(0)}}
\nc{\al}{\alpha}
\nc{\ga}{\gamma}
\nc{\de}{\delta}
\nc{\ep}{\epsilon}
\nc{\ze}{\zeta}
\nc{\et}{\eta}
\renewcommand{\th}{\theta}
\nc{\Th}{\Theta}
\nc{\ka}{\kappa}
\nc{\rh}{\rho}
\nc{\si}{\sigma}
\nc{\ta}{\tau}
\nc{\up}{\upsilon}
\nc{\ph}{\phi}
\nc{\ch}{\chi}
\nc{\ps}{\psi}
\nc{\om}{\omega}
\nc{\Ga}{\Gamma}
\nc{\De}{\Delta}
\nc{\La}{\Lambda}
\nc{\Si}{\Sigma}
\nc{\Up}{\Upsilon}
\nc{\Ph}{\Phi}
\nc{\Ps}{\Psi}
\nc{\Om}{\Omega}
\nc{\ptl}{\partial}
\nc{\del}{\nabla}
\nc{\ov}{\overline}
\nc{\newcaption}[1]{\centerline{\parbox{15cm}{\caption{#1}}}}
\nc{\none}{${\cal N}=1\;$}
\nc{\ntwo}{${\cal N}=2\;$}
\nc{\nfour}{${\cal N}=4\;$}
\nc{\nones}{${\cal N}=1^*\;$}
\nc{\Z}{\mathbb{Z}}

\nc{\C}{\mathbb{C}}

\begin{document}

\begin{titlepage}

\rightline{hep-th/0612077}

\setcounter{page}{1}

\vspace*{0.2in}

\begin{center}

\hspace*{-0.6cm}

\Large{\bf Superconformal $R$-charges and dyon multiplicities \\ in \boldmath{\ntwo}\!\! gauge theories}

\vspace*{0.5cm}
\normalsize

{\bf Adam Ritz}

\smallskip
\medskip

{\it Department of Physics and Astronomy, University of Victoria, \\
     Victoria, BC, V8P 5C2 Canada}

\smallskip
\end{center}
\vskip0.4in

\centerline{\large\bf Abstract}

${\cal N}=(2,2)$ theories in 1+1D exhibit a direct correspondence between the $R$-charges of chiral operators at a conformal point
and the multiplicities of BPS kinks in a massive deformation, as shown by Cecotti and Vafa. We obtain an analogous relation in 3+1D
for \ntwo gauge theories that are massive perturbations of Argyres-Douglas fixed points, utilizing the geometric engineering approach
to \ntwo vacua within IIB string theory. In this case the scaling dimensions of a certain subset of chiral operators at the UV fixed point
are related to the multiplicities of BPS dyons. When the Argyres-Douglas SCFT is realized at the root of a baryonic Higgs branch, this translation 
from 1+1D to 3+1D can be understood physically from the relation between the bulk dynamics and the  ${\cal N}=(2,2)$ worldsheet dynamics
of vortices in the baryonic Higgs phase. Under a relevant perturbation, the BPS kink multiplicity on the vortex worldsheet translates to 
that of the bulk dyonic states. The latter viewpoint suggests  the 3+1D version of the Cecotti-Vafa relation may hold more generally,  and
simple tests provide evidence in favor of this for more generic choices of the baryonic root.

\vfil
\leftline{December 2006}

\end{titlepage}

\renewcommand{\theequation}{\arabic{equation}}

\subsection*{1. Introduction}

The modern Wilsonian interpretation of quantum field theory, with its emphasis on conformal fixed points and the renormalization
group (RG) flows which connect them, naturally leads on to questions concerning the general classification of this RG structure.
In generic  quantum field theories in 3+1D, making progress in this direction appears a daunting problem, as in practice we still 
have relatively little knowledge of the basic configuration space, namely the space of allowed conformal fixed points. The addition of 
supersymmetry naturally assists in this regard, and recent developments such as $a$-maximization \cite{iw} have allowed
significant progress to be made in terms of understanding the properties of \none superconformal points, and some of the flows that
connect them \cite{iw2}. 

In the more restricted context of field theories in 1+1D, the situation is more clear cut. There are general schemes for
classifying fixed points and also powerful constraints, such as the Zamolodchikov $c$-theorem \cite{cthm},
which constrain the allowed RG flows
between them. The situation with ${\cal N}=(2,2)$ supersymmetric theories is under even greater control, to the extent that general classification
theorems \cite{cv} have been discussed for the full space of quantum field theories, conformal or otherwise. One aspect of this control is
the elucidation of some interesting relations between properties of  superconformal field theories (SCFTs) and certain features of the non-conformal 
theories to which the CFTs flow under relevant perturbations. In the context of  ${\cal N}=(2,2)$ theories, one of the
most interesting was obtained by Cecotti and Vafa \cite{cv}. It can be written in the form,
\be
 {\rm evals}({\cal M}) = \exp\left[2\pi i \left( q_k - \frac{\hat{c}}{2}\right)\right], \label{cv1}
\ee
where the monodromy matrix ${\cal M}$ on the left-hand side, to be discussed further below, is determined by the 
BPS soliton spectrum in the massive deformation
of a UV SCFT, while the right-hand side is given by the $R$-charges $q_k$ of operators in the chiral ring of  the undeformed SCFT, and the corresponding central 
charge $\hat{c}$. In this sense it is a nontrivial UV-IR relation, albeit a special one that relates protected BPS
quantities in a supersymmetric theory.

It is tantalizing to think that similar relations may exist in 3+1D. In general, given the significantly richer phase structure and
space of allowed relevant deformations, this seems a forlorn hope, but we will argue that in certain cases such constraints must hold, and 
indeed that the UV-IR relation retains essentially the same interpretation. The specific example we have in mind 
concerns \ntwo gauge theories that are massive deformations of specific Argyres-Douglas SCFTs \cite{ad}. In particular, 
we will argue that the simplest proof of (\ref{cv1}) in 1+1D, for Landau-Ginzburg theories \cite{cv}, can actually be translated quite directly
to the 3+1D theory by making use of the geometric engineering approach to \ntwo vacua in IIB string theory \cite{klmvw,ge2,lerche,sv}. The superpotential of the 
Landau-Ginzburg model emerges from the geometry of a local Calabi-Yau singularity, and the associated holomorphic structure required 
for this particular proof of (\ref{cv1}) in 1+1D can be translated  to the 3+1D \ntwo gauge theory realized at low
energy in this background.

The appearance of the analogue Landau-Ginzburg system in the construction outlined above  can be given a clear physical interpretation if this
SCFT is realized at the root of the baryonic Higgs branch in \ntwo SCQD with gauge group U($N$) and $N_f=N$ fundamental hypermultiplets,
In particular, in recent years a very precise correspondence  between the dynamics of these theories and the  ${\cal N}=(2,2)$ worldsheet dynamics of 
BPS vortices has been developed, where the vortices in question are present within the theory in the baryonic Higgs phase itself  \cite{ht1,abeky,sy,ht2,sakai}. 
In the interior of the vortex, a U(1) subgroup of the unbroken gauge symmetry at the root of the baryonic branch 
is restored, and it is in this sense that the vortices capture certain features of the full quantum gauge dynamics in the Coulomb phase of the 
theory. Indeed, beyond providing a precise window to the 
quantum vacuum structure on the Coulomb branch, there is also an exact map between a subset of BPS states; in particular confined monopoles in the 
bulk of the Higgs phase translate to BPS kinks in the worldsheet vortex dynamics \cite{mon}. In this context, if the baryonic root is tuned near to
an Argyres-Douglas point, it has recently been shown that the worldsheet dynamics of the vortex is a Landau-Ginzburg system, specifically a perturbed
minimal model \cite{tong}. This dynamics precisely realizes the analogue 1+1D system appearing in the geometric engineering construction.
Thus, this correspondence allows for a direct translation of (\ref{cv1}), interpreted now as a relation within the worldsheet theory of the vortex, to 
the physics of the bulk gauge theory in 3+1D at the root of the baryonic Higgs branch. 
In detail, this mapping applies only to the
special case where the baryonic root is a massive perturbation of an Argyres-Douglas SCFT, and does not immediately generalize to more generic
examples. Nonetheless, the existence of a more general proof of (\ref{cv1}) in 1+1D \cite{cv} and the general correspondence between vortex dynamics
and the bulk gauge theory,  suggests that the translation to 3+1D should have
more general validity. Indeed some exploratory checks appear to confirm that the relation does hold for more generic choices of the
baryonic root. 

We will begin in \S2 by first discussing the geometric engineering approach to \ntwo gauge theories, and specifically the regime near an Argyres-Douglas point
as considered by Shapere and Vafa \cite{sv}, which allows the 1+1D proof of (\ref{cv1}) to be lifted to the bulk theory via the identification of an analogue 
Landau-Ginzburg superpotential in  the complex structure of the geometry. In \S3, we turn to the interpretation of this mapping from 1+1D to 3+1D,
and describe how the dynamics of vortices in SQCD, recently discussed by Tong \cite{tong}, provides a physical realization of the analogue Landau-Ginzburg
system. This point of view also suggests that (\ref{cv1}) may have more general validity in 3+1D and we find that anecdotally it does hold for a more general
choice of the baryonic root, where both the bulk and worldsheet physics is somewhat different. We summarize in \S4 with some additional 
remarks and possible generalizations.

\subsection*{2. Perturbed Argyres-Douglas points and BPS states}

In testing a relation such as (\ref{cv1}) in 3+1D, we require first of all a convenient technique to determine the multiplet structure
of BPS states. In \ntwo gauge theories, computing the degeneracies of  BPS monopoles and dyons is a rather nontrivial problem in general,
and has not been studied in detail away from the weak coupling regime, with the exception of a few special cases with, for example,  an SU(2)
gauge group.  Fortunately, we are concerned here only with those states that become massless at the conformal point, which is therefore a subset 
of the full BPS spectrum, that which is massive, but light, in a region of the Coulomb branch near the Argyres-Douglas point.
The geometric  engineering approach  to \ntwo theories within IIB string theory provides a very convenient framework for this 
analysis \cite{sv} as it maps the problem directly to the analogous one of counting BPS states in an analogue 1+1D theory. 

The basic geometry in this case is a noncompact Calabi-Yau 3-fold specified by the hypersurface $P(x_i)=0$ in $\C^4$, where
\be
 P = F(x,y) + z^2 + w^2,
\ee
and $F(x,y)$ is a quasi-homogeneous function of $x$ and $y$ describing a Riemann surface $\Si$, which
is just the Seiberg-Witten curve of the gauge theory \cite{sw},
\be
 \Si: F(x,y)=0.
\ee
The holomorphic 3-form is given by 
\be
 \Om = \frac{\prod_i dx_i}{dP},
\ee
which determines the mass of particles in the effective low-energy gauge theory realized as $D$-branes wrapping cycles in the 
geometry. In particular for D3-branes wrapping supersymmetric 3-cycles $C$, one finds BPS states with central charge
\be
 {\cal Z} = \int_C \Om. \label{mass}
\ee

We will be interested in Calabi-Yau 3-folds with singularities which realize a special class of perturbed Argyres-Douglas points in 
the gauge theory. To this end we can specify
\be
 F(x,y) = y^2 + ({\cal W}'(x))^2, \label{swcurve}
\ee
where ${\cal W}$ is a polynomial of degree $N+1$,
  \be
  {\cal W}(x,g_j) =\frac{1}{N+1} x^{N+1} + \sum_{j=2}^{N} \frac{g_j}{N+1-j} x^{N+1-j}. \label{Wp}
 \ee 
 The interacting superconformal point arises in the infrared when $g_j=0$, for all $j$, where additional cycles vanish in the geometry
leading to a certain subset of the BPS dyon states,\footnote{\,These states in general carry topological and global charges, and 
we will refer to them generically as ``dyons''.} realized as wrapped D3-branes, becoming massless at this point. 
 
 \subsubsection*{2.1 \boldmath{$R$}-charges and the chiral ring}

The existence of mutually nonlocal massless degrees of freedom at the conformal point implies that there is no straightforward local 
Lagrangian description and the Argyres-Douglas fixed points, despite their extended \ntwo supersymmetry, are still quite poorly
understood. Fortunately, for our purposes it will be sufficient to know the scaling dimensions of the chiral primary operators dual to
the perturbations $g_j$. This
information is BPS-protected and fixed by the chiral ring relations, which are determined by the local structure of the geometry near the
conformal point. Indeed, for the present case, this data was determined in \cite{eh,sv} using the approach pioneered in
\cite{apsw}. The basic idea is that in the infrared SCFT a U(1)$_R$ symmetry must necessarily be restored, and this symmetry must be
visible in the Seiberg-Witten curve. The chiral primaries corresponding to deformations along the Coulomb branch have 
vanishing SU(2)$_R$ spin and so have dimension $D=R/2$ in terms of the conformal U(1)$_R$ charge. 
To retain  the required U(1) scaling symmetry of the curve which is present at the 
conformal points, where $D[y]=N D[x]$, one observes from (\ref{Wp}) that we require $D[g_j]=j D[x]$, while $D[x]$ can be 
normalized using the fact that $D[\Om]=1$ from (\ref{mass}). One obtains \cite{sv},
\be
 D[g_j] = \frac{j}{N+1}, \;\; j = 2,\ldots N. \label{Dg}
\ee
The chiral operators ${\cal O}_j$ dual to these perturbations then have $D[{\cal O}_{j}]=2-D[g_j]$ \cite{apsw,sv}
as follows from the dimension of the superpotential.

\subsubsection*{2.2 BPS states and the Cecotti-Vafa relation}

The relevant BPS states in this geometry were discussed in \cite{sv}, and are given by D3-branes wrapping supersymmetric 3-cycles.  One 
constructs the required cycles as 2-spheres,
a real subspace of $y^2+z^2+w^2=-({\cal W}'(x))^2$, fibered over a real curve in the $x$-plane \cite{klmvw}. The latter curve begins and ends at
extremal points of ${\cal W}(x)$ where the radius of the $S^2$ goes to zero. A possible basis of 3-cycles is then given by choosing the
$2N-1$ intervals between the zeros of $({\cal W}'(x))^2$ in a sequence. The intersection of the 3-cycles then gives the Dynkin diagram
of $A_{2N-1}$. For these cycles to be supersymmetric, there is the additional condition that along a given path between two zeros
of ${\cal W}'(x)$ the phase of $\int_{S^2} \Om$ should be constant \cite{sv}, which ensures that the BPS mass inequality is saturated.
This reduced one-form on the $x$-plane is precisely the Seiberg-Witten differential, and in general takes the form \cite{klmvw},
\be
 \lambda_{SW} = \int_{S^2} \Om = x \frac{dt}{t}, \label{lam}
\ee
where $2t=y+ ({\cal W}'(x)+M)$, with $M$ a dimensionful constant, that can be interpreted as setting a fixed scale in the geometry that will be used shortly to
consider a scaling limit near the singularity. Denoting the zeros of ${\cal W}'(x)$ as $x=e_i$, for $i=1,\ldots,N$, the central charge 
which determines the mass of the BPS states takes the form,
\be 
  {\cal Z}_{ij} = \int_{e_i}^{e_j} x\frac{dt}{t} =  \int_{e_i}^{e_j} \frac{x{\cal W}''(x) }{{\cal W}'(x)+M} dx.
  \ee
 
 We can now go further and take a scaling limit near the Argyres-Douglas point, corresponding to $x/M \rightarrow 0$, and after a constant rescaling
 of $x$ and $g_i$ we find,
 \be
  {\cal Z}_{ij}  = \int_{e_i}^{e_j} \lambda_{SW} \longrightarrow  \int_{e_i}^{e_j} {\cal W}'(x) dx = {\cal W}(e_j) - {\cal W}(e_i). \label{phase}
 \ee
 We conclude that the constant-phase condition for the integrand in (\ref{phase}) can be restated as $\int_{x(0)}^{x(t)} d{\cal W}(x) = \al t$, where 
$\al$ is a constant and $t\in [0,1]$ parametrizes the real path
between the zeros  of ${\cal W}'(x)$. In other words ${\cal W}(x)$ must follow a 
straight line between  two extrema in the $x$-plane,
\be
 {\cal W}(x(t)) = {\cal W}(e_i)+ t ({\cal W}(e_j) -{\cal W}(e_i)),
\ee
which we recognize as precisely the condition for a BPS kink trajectory within the Landau-Ginzburg model with superpotential
${\cal W}(x)$ \cite{fmvw}, which in this case is a perturbation of the $A_{N-1}$ minimal model.

At this point it becomes clear that counting BPS solitons in this setup is entirely equivalent to the problem of
counting BPS kinks in the perturbed minimal model,\footnote{\,This correspondence was noted in passing in \cite{sv}.} and more 
precisely the approach used by Cecotti and Vafa \cite{cv}  for performing the latter calculation translates directly to this theory. 
The answer in both cases is therefore the same and it is rather remarkable that the same technique may be used to perform the calculation, at 
least in this special example. It is worth emphasizing that the manipulations of the Seiberg-Witten differential leading to 
(\ref{Ws}) are rather  particular to this case, and would not apply to a more generic point on the Coulomb 
branch.\footnote{\,The simplification of $\lambda_{SW}$ in this limit near the Argyres-Douglas point appears related to a recent discussion
of Strebel differentials in \ntwo gauge theories \cite{cachazo}.}

The answer one obtains for the BPS multiplicity in fact depends on the
precise form of the deformation parameters $g_i$, due to the presence of marginal stability curves across which the BPS spectrum can jump. To briefly
summarize the results \cite{cv}, note that near each critical point ${\cal W}(e_i)$ one can define a vanishing cycle $\De_i$ in the space of fields,
specified by a fixed value of ${\cal W}(t)$. In the Landau-Ginzburg system, this is the set of all possible solutions to the linearized Bogomol'nyi equation 
about the vacuum ${\cal W}(e_i)$, but naturally arises here as the set of vanishing cycles associated with the singularity in the full Calabi-Yau geometry.

As shown in \cite{cv}, the CFIV index \cite{cfiv} counting
the number of BPS multiplets between two vacua is given by
\be
 \mu_{ij} = \De_i \circ \De_j,
\ee
where the product is the intersection number of the two cycles. An important aspect of the arguments demonstrating this relation
in \cite{cv} is that the changes in the BPS spectrum that occur on crossing a marginal stability curve translates directly into 
the Picard-Lefschetz monodromy that the intersection number can undergo with changes in parameters. The index also carries
a phase associated in 1+1D with the (fractional) fermion number of the state, $\exp(2\pi i (f_{i}-f_{j}))$, where in 
the present case \cite{fi,cfiv},
\be
 f_j = -\frac{1}{2\pi}{\rm Arg}({\cal W}''(e_j)). \label{fermi}
\ee
In particular, since the K\"ahler metric is real,  $f_j$ is just the phase of the fermion mass (more generally the phase of the determinant of the mass matrix) 
in the $j$-vacuum. It is natural to suspect a similar interpretation in 3+1D, where dyon solutions do in general exhibit charge fractionalization
at generic points on the Coulomb branch \cite{weffect}. We will return to this point again below.

From the index, we can build the monodromy matrix in 3+1D, 
\be
 {\cal M}_{\rm 4D} = S^{-T}S\;\;\;\;\; {\rm with}\;\;\;\;\; S_{ij} = \de_{ij} - \mu_{ij}, \;\; {\rm for}\;\;  i\leq j, \label{monod}
 \ee
 given a deformation leading to a convex vacuum polygon in the ${\cal W}$-plane. It is the eigenvalues of this
matrix which provide the link between the UV and IR properties of the theory. From the discussion above it
is clear that these eigenvalues are determined on the one hand by the index counting BPS states in the 
massive deformation. 
 
 On the other hand, the ensuing topological stability of the index allows for an alternative derivation of the eigenvalues of the monodromy matrix,
by collapsing all the critical values of ${\cal W}$ to a point. i.e turning off all the mass deformations. In 
this conformal limit ${\cal W}$ is quasi-homogeneous and the
eigenvalues of ${\cal M}$ can be obtained from a basis of forms dual to the space of vanishing cycles $\De_i$, as described
in \cite{cv}. In the present case,
where ${\cal W}$ is a function of just one holomorphic variable, we have the set of 1-forms $\om_k=\ph_k dx$, where $\ph_k$ is a monomial basis
for the chiral ring with $R$-charge (or equivalently degree)  $q_k$. Up to changes of basis, these $R$-charges are equivalent to (twice) the
scaling dimensions in (\ref{Dg}), as the latter were inferred directly from ${\cal W}$ in the same way.
Writing $\om_k =\ph_k dx = \al_k  dW$,  which is well-defined away from the critical 
points, the functions $\al_k$ form a basis for the dual-space to the cycles $\De_i$. Now on a circular path around a given vacuum
${\cal W}(e_k)$ in the ${\cal W}$-plane, $x$ rotates by a phase of $2\pi q_k$. Consequently we may read off the phase by which $\al_k$ rotates
which gives directly the corresponding eigenvalue of ${\cal M}$ \cite{cv}. 

The result of equating both calculations is the Cecotti-Vafa relation \cite{cv}, 
\be
 {\rm evals}({\cal M}_{\rm 4D}) = \exp\left[2\pi i \left( q_k - \frac{c'}{2}\right)\right], \label{cv2}
\ee
where in this case $c'=1-2/(N+1)$. Our primary observation here is that, since the derivation is purely geometrical,
its interpretation translates directly from the analogue 1+1D system with superpotential ${\cal W}$ to 3+1D,  given the appropriate
identification of the BPS states. We have just described how the data on the left-hand side, namely the 
BPS spectra of dyons, maps directly from that of the kink spectrum in a massive deformation of the $A_{N-1}$ model in 1+1D. 
In turn, the spectrum of $UV$ $R$-charges determined directly from the geometry of the CY-fold near the 
Argyres-Douglas point is specified by the superpotential ${\cal W}$ and 
consequently by the chiral ring of the 1+1D model. However, an important distinction is that we have only considered a subset of all possible 
deformations of the Seiberg-Witten curve. Due to the identification in (\ref{swcurve}), we are only considering half the allowed deformations, namely those
that do not disturb the factorized form of the curve, and identifiable with `normalizable' (i.e. mass) perturbations \cite{sv}, which therefore keep the vevs fixed. 
We should also note that while $c'$ can be identified with the central charge of the $A_{N-1}$ minimal model, it is not 
the full central charge of the string theory on the CY, as the 1+1D analogue only captures certain parts of the geometry, and the complex dimensions parametrized
by $z$ and $w$ decouple.  This reflects the fact that we are not considering all possible BPS states here but only those that become
light specifically at the conformal point. In the next section, we will provide a physical rationale for the appearance of this analogue 1+1D theory within this construction.

\subsection*{3. Interpretation via the vortex worldsheet and generalizations}

The direct translation of the Cecotti-Vafa relation from 1+1D to 3+1D discussed above can actually be given a natural physical 
interpretation if we realize the Argyres-Douglas SCFT at the root of a baryonic Higgs branch. The 1+1D theory that appears as 
an analogue system in the previous section then arises directly, describing the worldsheet dynamics of BPS vortices which
are present in the baryonic phase. This viewpoint also implies that the Cecotti-Vafa relation may have more general validity in
3+1D beyond the special case discussed in \S2. In this section, we will expand on this point of view and also provide some anecdotal
evidence supporting the general validity of the Cecotti-Vafa relation in theories of this type.

\subsubsection*{3.1 Argyres-Douglas SCFT at the baryonic root}

\ntwo SQCD with gauge group U($N$) and $N_f=N$ fundamental hypermultiplets $(Q,\tilde{Q})$ exhibits a rather complex vacuum 
structure, comprising a Coulomb branch parametrized, for example, by the vev of the adjoint scalar $\Ph$ in the vector multiplet, with components
$\ph_a$, $a=1,\ldots,N$, and various Higgs branches parametrized by the vevs of the squark multiplets $Q_i^a$ and $\tilde{Q}_a^i$. The Higgs
branches connect to the Coulomb branch at special points determined by the mass parameters $m_i$ of the squarks. The vacuum branch
relevant here is the baryonic branch parametrized by a gauge invariant baryon vev, $B = \ep_{a_1..a_N} Q^{a_1}_1 \ldots Q^{a_N}_N$. 
This baryonic  branch connects to the Coulomb branch at the baryonic root, given classically by $\ph_a=m_a$, which generically is characterized 
by the presence of additional massless quark hypermultiplets, but is not otherwise special. However, if the mass parameters $m_a$ are tuned appropriately, then at the 
quantum level one finds that additional monopole multiplets also become massless at this point. The mutually nonlocal massless degrees of freedom
signify an interacting Argyres-Douglas SCFT.  A schematic outline of this subset of the moduli space is shown in Fig.~1.

\begin{figure}
\begin{picture}(450,100)
\put(0,0){\centerline{\includegraphics[width=9cm]{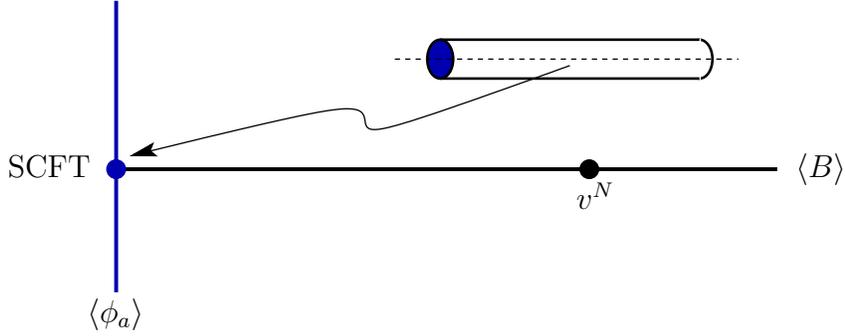}}}
\put(100,-10){{$\langle \ph_a\rangle$}}
\put(368,45){{$\langle B \rangle$}}
\put(70,45){{SCFT}}
\put(285,33){{$v^N$}}
\end{picture}
\vspace*{0.1in}
\caption{\footnotesize  A schematic view of the vacuum structure, comprising a slice through the Coulomb branch parametrized by 
$\ph_a$, $a=1,\ldots,N$,
which for $\ph_a=m_a+{\cal O}(\La)$ connects to a baryonic Higgs branch parametrized by the baryon vev $B$, and the classical vacuum
at $\langle B\rangle = v^N$ admits BPS vortex solutions. The vacuum probed by the interior of a given vortex is a point on the 
Coulomb branch at the baryonic root and at the quantum level, for a suitable choice of $m_a\sim {\cal O}(\La)$, the baryonic root can 
be tuned to an Argyres-Douglas SCFT. }
\end{figure}

To make contact with the discussion in \S2, recall that for $N_f=N$ the Seiberg-Witten curve takes the form \cite{ho,aps1,witten},
\be
 y^2= \prod_{a=1}^N (x-\ph_a)^2 - 4\La^N \prod_{k=1}^N (x-m_k). \label{adcurve}
\ee
Specializing  to the baryonic  root, where $\prod_a(x-\ph_a) = \prod_k(x-m_k) +\La^N$ \cite{aps,dorey,dht}, this curve degenerates 
and one finds that $N$ quark hypermultiplets become massless, allowing for the baryon vev to be turned on.  The degenerate
curve can be written as,
\be
 y^2= ({\cal W}'(x))^2,\;\;\;\;{\rm with}\;\;\; {\cal W}'=\prod_{k=1}^N (x-m_k) - \La^N, \label{W2}
  \ee
 which is precisely of the form (\ref{Wp}),  with the deformation parameters $g_i$ re-expressed in terms of the
 hypermultiplet masses $m_k$, such that $g_1=\sum_k m_k=0$. Moreover, as recently discussed in \cite{tong}, further 
 tuning to the special  point $g_{j}=0$, $j=2,\ldots,N$ (i.e. $m_k=-\exp(2\pi i k/N) \La$ for $N$-odd and $m_k=-\exp(2\pi i (k+1/2)/N)\La$ for $N$-even), 
 leads in addition to a fully degenerate curve $y^2=x^{2N}$ and an interacting SCFT in the infrared.
 
 Having realized the conformal point, and small deformations thereof, in this setting we can now make use of the known correspondence 
 between the worldsheet dynamics of vortices on the baryonic branch, where  $\langle B\rangle = v^N$ with $v$ a Fayet-Iliopoulos term,
 and the bulk dynamics at the baryonic root \cite{ht1,abeky,sy,ht2,tong} to reinterpret the results of \S2. As shown schematically in Fig.~1, while 
 vortices exist only in the theory
 on the baryonic branch for $v\neq 0$, in the center of the vortex a U(1) subgroup of the Coulomb phase gauge symmetry at the baryonic root is restored. It is in this sense 
 that the vortex acts as a probe of the theory at special points in the Coulomb phase \cite{tong}, i.e. in this case at an Argyres-Douglas SCFT.

\subsubsection*{3.2 The \boldmath{$A_{N-1}$} worldsheet theory}

Vortices living on the baryonic Higgs branch are known to inherit a nontrivial dynamical structure on the worldsheet from the fact that they
spontaneously break some of the residual flavor symmetry of the theory. In the simplest scenario relevant to the above discussion,
this endows the vortex with a set of bosonic moduli parametrizing the coset $\C P^{N-1} = $U($N$)/(U($N-1)\times$ U(1)) interpreted
as Goldstone modes \cite{ht1,abeky}. Since these states are BPS within the \ntwo theory, this bosonic structure is completed to a  ${\cal N}=(2,2)$ supersymmetric
sigma model. There are in addition further modes associated with the breaking of translational invariance in the directions transverse
to the vortex, but these will play no role here.

At the quantum level, the effective superpotential $\widetilde{\cal W}$ for the ${\cal N}=(2,2)$ $\C P^{N-1}$ theory can be written in terms of a 
twisted scalar $\Si$ and twisted masses $\tilde{m}_k$, where\footnote{\,The branch structure in this expression reflects shifts in the $\th$-parameter (giving the
phase of $\tilde{\La}$), and in 1+1D the ambiguity is fixed by minimizing the electric potential associated with $\th$ when moving to this effective description 
\cite{coleman,witten93,hh,dorey}.}
\be 
 \frac{\ptl \widetilde{\cal W}}{\ptl \Si} = \ln \prod_{k=1}^N \left(\frac{\Si - \tilde{m}_k}{\tilde{\La}}\right), \label{Weff}
\ee
so that the vacuum constraint $\widetilde{\cal W}'(\Si)=0$, leading generically to $N$ massive vacua, matches the singular point  ${\cal W}'(x)=0$ for the bulk curve (\ref{W2}), 
given the identifications $x \leftrightarrow \Si$, $\La \leftrightarrow \tilde{\La}$ and $m_k \leftrightarrow  \tilde{m}_k$, between the chiral fields and mass scales. 
Thus the worldsheet dynamics captures the vacuum structure at the baryonic root,
as specified by the Seiberg-Witten curve, and indeed the central charge for (a subset of) the bulk dyon states can be written in a form
precisely matching that for the BPS kinks which interpolate between the $N$ vacua of the worldsheet theory  \cite{dorey},
\be 
 {\cal Z}_{ij} = \int_{e_i}^{e_j} \widetilde{\cal W}'(x) dx = \widetilde{\cal W}(e_j) - \widetilde{\cal W}(e_i), \label{cc}
 \ee
where $e_i$ are the zeros of ${\cal W}'$ and thus also of $\widetilde{\cal W}'$. 
 
In general the functions ${\cal W}$ and $\widetilde{\cal W}$ are not directly
related, except through the structure of their extremal points.
However, a more precise correspondence does arise if we take a scaling limit and tune the baryonic root close to an 
Argyres-Douglas point. Indeed, the $N$ massive vacua of this theory all merge at a critical point $\langle \Si \rangle =0$ when the 
twisted mass parameters $\tilde{m}_k$,  identified with the quark mass parameters in the bulk theory $\tilde{m_k}=m_k$, are
set to the conformal point \cite{svz,tong}. Generically, the theory has BPS kinks interpolating between the $N$ vacua, and 
these all become massless at this special point, where
\be 
 \left.\frac{\ptl \widetilde{\cal W}}{\ptl \Si}\right|_{\rm crit} =  \ln \left(1+\frac{\Si^N}{\tilde{\La}^N}\right) \longrightarrow  
 \frac{\Si^N}{\tilde{\La}^N} + \cdots .
 \ee
Consequently, as recently observed by Tong \cite{tong}, taking the scaling limit near the critical point $\Si/\tilde{\La} \rightarrow 0$ and retaining 
only the leading term (in analogy with the corresponding limit taken in (\ref{phase})), one finds on rescaling $\Si$  and parametrizing the 
relevant mass deformations in terms of $\tilde{\nu}_j$,
\be
 \widetilde{\cal W}(\Si) \longrightarrow {\cal W}(x=\Si,\tilde{\nu}_j) \sim \frac{\Si^{N+1}}{N+1} + \sum_{j=2}^N \frac{\tilde{\nu}_j}{N+1-j} \Si^{N-j+1}+ \cdots,\label{Ws}
 \ee
 which exactly matches the perturbed bulk curve and the analogue superpotential in (\ref{Wp}). For $\tilde{\nu}_j=0$ this 
 is the superpotential of a theory which flows in the infrared to the
  ${\cal N}=(2,2)$ $A_{N-1}$ minimal model with normalized central charge $\hat{c} = 1-2/(N+1)$ \cite{witten1}. 
   Since $D[\Si]=1/(N+1)$ at this point, as follows from (\ref{Ws}), we may read off the  dimensions of the 
 perturbations $\tilde{\nu}_j$ finding $D[\tilde{\nu}_j] = j/(N+1)$ in precise agreement with the bulk SCFT \cite{tong}, while from (\ref{cc}) we also see that
 the central charges exactly match the bulk prediction (\ref{phase}) in this limit. Note that the monodromy structure for BPS states in $\C P^{N-1}$ with twisted masses
 \cite{hh,dorey} has been truncated in this limit, as the central charge (\ref{cc}) will now exhibit only a finite number of branches in the parameter space $\{\tilde{\nu}_j\}$.
 However, this is precisely the truncation we would expect on limiting our attention to the states which are light in the vicinity of the critical point. 
 We conclude that, in this particular case, we can make a precise identification
 between the analogue 1+1D Landau-Ginzburg theory emerging from the geometry in \S2 and the worldsheet dynamics of vortices in the baryonic Higgs phase.
 
 This point of view also suggests that the limiting procedure implicit in the right-arrow symbol in (\ref{Ws}) (and thus also in (\ref{phase})) implies a 
 precise order of operations, namely that one first tunes to the $A_{N-1}$ critical point and then perturbs by turning on $\tilde{\nu}_j$; it is not clear 
 that the same RG trajectory is recovered if these operations are performed in the reverse order. Indeed the $\C P^{N-1}$ model is not a relevant perturbation of this CFT, as 
 there are additional irrelevant operators that have been dropped  in (\ref{Ws}). Thus the perturbations of the critical point, introduced in this limit, need not be related
 trivially to the original twisted mass perturbations of the $\C P^{N-1}$ model.

For completeness, we can also consider the analogous limit in the mirror description of the $\C P^{N-1}$ model, namely the $\hat{A}_{N-1}$ affine-Toda theory
with superpotential \cite{hv}, 
\be
 \widetilde{\cal W}_{M}(X_i) = \tilde{\La} \left( \sum_{i=1}^{N-1} e^{X_i} + \prod_{i=1}^{N-1} e^{-X_i} \right) + \sum_{i=1}^{N-1} (\tilde{m}_i - \tilde{m}_N) X_i.
\ee
The conformal point arises for the same choice of twisted masses, where the vacuum takes the form $\langle X_k \rangle=2\pi ik/N$, and in this case one
may verify directly that only a single massless state arises (a point that was an implicit assumption above in assuming the validity of (\ref{Weff})). We can write
the effective superpotential for this light mode, $\widetilde{\Si}$, by perturbing near the conformal point, $e^{X_i} = \langle e^{X_i}\rangle + \widetilde{\Si}/\La$, 
obtaining (up to a constant),
\be
 \widetilde{\cal W}_{M}(\widetilde{\Si}) \longrightarrow - \frac{1}{N+1} \frac{\widetilde{\Si}^{N+1}}{\tilde{\La}^N} + \cdots,
 \ee
 consistent with the conformal limit of the $\C P^{N-1}$ superpotential. This is in agreement with the fact that the $A_{N-1}$ minimal model
 is self-mirror, up to a trivial $\Z_{N+1}$ orbifold action \cite{hv}, and provides a nice consistency check on the validity of the effective description
 near the conformal point.

\subsubsection*{3.3 Generalization to the generic baryonic root}

With this physical viewpoint in mind, identifying the analogue 1+1D model 
with the worldsheet theory on vortices present in the Higgs phase, its natural to enquire whether the correspondence
may be more far-reaching. In particular, Cecotti and Vafa were able to prove the relation (\ref{cv1}) in a far wider class of
theories than just those with a Landau-Ginzburg realization, and it seems reasonable to ask whether it similarly has a greater scope in 
3+1D. One obvious candidate concerns a more generic parameter choice for the baryonic root in the $N_f=N$ theory, where the 
correspondence between bulk and vortex worldsheet dynamics implies a direct relation between the BPS spectra. We will now
comment briefly on the possibility that the Cecotti-Vafa relation also extends to 3+1D in this more generic case.

As discussed above, the
worldsheet theory for a generic choice of the baryonic root is an  ${\cal N}=(2,2)$ sigma model with target space $\C P^{N-1}$, arising from the broken flavor
symmetries. If we consider the simplest case with twisted mass parameters set to zero, namely a point well inside the marginal stability curve discussed
in \cite{dorey,svz}, the theory has $N$ massive vacua at the 
quantum level, and the BPS spectrum comprises kinks which interpolate
between these vacua. The degeneracies of these 1/2-BPS multiplets are known and, for kinks interpolating between vacua 
which differ in phase by $2\pi k/N$ units, are given by 
\be
 \mu_{ii+k} = (-1)^{k-1} \left(\begin{array}{c} N \\ k \end{array}\right) \;\;\;\;\;
  \Longrightarrow \;\;\;\;\; {\rm evals}({\cal M}_{\rm 2D}) = (-1)^{N-1},\label{cpnin}
  \ee
where ${\cal M}_{\rm 2D}$ is the corresponding monodromy matrix in 1+1D defined as in (\ref{monod}). The eigenvalues are then $N$-fold degenerate.

The UV $R$-charges of generators of the chiral ring are known in this case to be given by the dimensions of the harmonic
forms of $\C P^{N-1}$ \cite{cv},
 \be
  q_k = k, \;\;\;\; {\rm for} \;\;\;\; k=0,\ldots ,N-1.
 \ee
 and we can write ${\rm evals}({\cal M}_2)= \exp(2\pi i (q_k - \hat{c}/2))$ where $\hat{c} =(N-1)$ is the normalized central charge.
 This exhibits one of the standard examples of the Cecotti-Vafa relation for sigma models with $\Z_N$ symmetry.
 
 Moving to the bulk theory, the BPS kinks are interpreted in the Higgs phase as monopoles, and so we expect the spectra of these two
 sets of states to match. This has indeed been verified at the semiclassical level at the root of the baryonic Higgs branch \cite{dht}. At this
 point, the relevant subset of the BPS monopole spectrum is given by a set of 
 bound states of the fundamental monopole with the $N$ quark flavors. At the classical level, this is visible as a set of $N$ fermionic flavor 
 zero modes inherited by the monopole solution so that, within semiclassical quantization, if the `bare' monopole solution is
 denoted $|0\rangle$, the bound states are obtained by acting with the appropriate fermionic creation operators, corresponding to
 each fermionic zero mode, $\rh^\dagger_{\lambda_1}\cdots \rh^\dagger_{\lambda_k} |0\rangle$. Although this analysis is not strictly
 valid when the quark mass parameters are set to zero, if we extrapolate to this case we find multiplets that 
  lie in the $k^{th}$ antisymmetric representation of the SU$(N)$ flavor symmetry, leading precisely to the result (\ref{cpnin}) for the multiplicity of 
  BPS states in the sector with $k$ flavor zero modes, with the same result for the monodromy matrix ${\cal M}_{4D}$.
 Nonetheless, one should bear in mind that the above analysis is semiclassical and, although the index is a 
 BPS-protected quantity, could be invalidated at the quantum level through the 
 presence of marginal stability 
 curves. However, the fact that the result matches the expectation (\ref{cpnin}) in this case suggests that no restructuring of this
 kind occurs for this particular choice of parameters.
  
 We may therefore ask whether the ensuing prediction for the spectrum of $R$-charges,  $q_k = k$ for $k=1,\ldots ,N-1$,
 for operators in the chiral ring matches with expectations. The UV limit  is simply the classical limit of the asymptotically
 free theory, where the $R$-symmetry is restored.  The generators of the chiral ring are given by 
 \ntwo completions of tr$(\Ph^k)$, tr$(\tilde{Q}\Ph^k Q)$, and baryonic operators in the case with $N_f=N$ that 
 we consider here \cite{chiral1,chiral2}. In the classical limit we will ignore the 
 operators associated with the gauge multiplet, tr$(W_\al W^\al \Ph^k)$, which are in any case presumably 
 part of the \ntwo completion of tr$(\Ph^k)$. Furthermore, since we are interested only in those deformations which keep the theory at
 the root of the baryonic Higgs branch, as determined by the classical superpotential ${\cal W} = {\rm tr}[\tilde{Q}(\Ph -  m)Q]$, we
 can limit our attention to the generators tr$(\Ph^k)$. Note that mesonic and baryonic perturbations affect only the mass and localization
 scale of the vortex in this case.
Limiting our attention to this reduced subset of the chiral ring, we have the kinematic constraint, rank$(\Ph) \leq N$, and 
so we can take a basis of such operators to be,
\be
 \{{\rm tr}(\Ph^1), {\rm tr}(\Ph^2),\ldots, {\rm tr}(\Ph^{N})\}.
\ee
In the UV limit, with SU(2)$_R$ spin equal to zero, these operators have U(1)$_R$ charge $Q_j=[{\rm tr}(\Ph^{j})]_R=2j$, for $j=1,\ldots,N$. 
Interpreting $Q_j=q_j+\tilde{q}_j$ in terms of the (1+1D) left- and right- chiral  U(1) charges \cite{gkp} in the  ${\cal N}=(2,2)$ 
superalgebra (with $q_j=\tilde{q}_j$ for the nonchiral $\C P^{N-1}$ model), we see that this exactly matches the spectrum
of $R$-charges for the UV chiral-ring generators in the $\C P^{N-1}$ model.

The arguments of the preceding section, providing a more direct link between the BPS structure and chiral rings in 1+1D and 3+1D do
not extend straightforwardly to this example. For a start, the $\C P^{N-1}$ model is not of Landau-Ginzburg type (even in its mirror 
formulation as an affine-Toda theory, where the superpotential goes to zero in the UV), and so one must rely on the more general proof
by Cecotti and Vafa of the UV-IR correspondence on the vortex worldsheet \cite{cv}. Given the apparent success of this anecdotal test of 
the Cecotti-Vafa relation in the
more general setting discussed above,  it would be interesting to see if this worldsheet structure also maps in some way to 
the bulk 3+1D theory.

 \subsection*{4. Concluding remarks}
 
 We have discussed a particular UV-IR relation in the context of RG flows in \ntwo gauge theories, and more specifically gave
 arguments that the Cecotti-Vafa relation between conformal scaling dimensions and BPS soliton multiplicities in 1+1D  translates directly to an 
 analogous correspondence in 3+1D. An important aspect of this was the role played by vortex dynamics in the gauge theory, which seems
 to provide the link between the natural 1+1D setting for the Cecotti-Vafa relation and its four-dimensional realization. We will conclude
 in this section with a few remarks on further aspects of the correspondence and possible generalizations. 
  
  One point touched on briefly in passing concerns the fractional fermion number, which appears as the phase of the CFIV index counting 
BPS states in 1+1D, and it is natural to ask whether it has a similar interpretation in the bulk.  The natural point of contact is
the fractional quark charge of dyonic states, given by  $2\pi \De S_i = {\rm Im} (\ptl {\cal Z}_{kl}/\ptl m_i)$ for the $i^{th}$ flavor \cite{fraction1,fraction2,fraction3},
which is analogous to the Witten effect shifting the electric charge \cite{weffect}, and indeed is equivalent at the baryonic root. For a given
extremal point $e_k$, we have the comparison,
\be
  S_i(e_k) = -\frac{1}{2\pi} {\rm Arg}\left[\frac{1}{e_k-m_i}\right] \; \longleftrightarrow \; f_k = -\frac{1}{2\pi} {\rm Arg}({\cal W}''(e_k)) = -\frac{1}{2\pi} {\rm Arg} 
   \left[\sum_i \frac{1}{e_k-m_i}\right], \label{fn}
 \ee
 suggesting that the worldsheet fermion number (\ref{fermi}) is identified with an appropriately defined sum over the fractional 
 quark charges.
One should bear in mind
that the quark states which become massless at the baryonic root are not visible on the worldsheet and these charges
must be decoupled in a relation such as (\ref{fn}). Consequently, in the simplest case with $N=2$ where there is just a single quark charge $S$, we find in 
the semiclassical limit for real $m=m_1-m_2$, $ \De f = \De S = \pm 1/2$, in agreement with \cite{svz,dorey} reproducing the half integer fermion number 
of both fundamental monopoles and $\C P^1$ kinks.

This viewpoint suggests a possible reinterpretation of charge fractionalization in 3+1D. In particular,  
on introducing spatial boundaries in 1+1D, one can view the fermion fractionalization of kinks as arising from  part of the integral 
fermion number residing at the boundaries \cite{svv}. This has a physical bulk interpretation in this case.\footnote{\,I'd like to thank A. Vainshtein
for helpful discussions on this point.} In particular, spatial boundaries on the vortex worldsheet arise directly if we consider a configuration
on the Higgs branch where the vortex lies between two domain walls \cite{sydw}. The dyon  central charge is independent of the
Fayet-Iliopoulos term and thus this effective dimensional reduction of the magnetic flux can be achieved without affecting charge fractionalization
which depends only on quantum corrections to the central charge. Consequently, we might expect the residual non-integer fermion number to
reside on the bounding domain walls in this case, and it would be interesting to explore this further.

Concerning the generality of relations such as (\ref{cv1}) and (\ref{fn}), the initial motivation for this study was the question of whether the 
Cecotti-Vafa relation had any analogues in the context of
\none gauge theories in 3+1D \cite{iw2}, where the structure of SCFTs has recently been elucidated in some detail using new techniques such 
as $a$-maximization \cite{iw} to pin down the spectrum of $R$-charges and scaling dimensions. While it is clear that such a correspondence
must be somewhat different to the \ntwo examples discussed here, conjecturally  relating chiral charges to BPS domain wall multiplicities, 
it would be interesting to explore whether soft breaking to \none can shed further light on this.

In this regard, it is also natural to wonder about a direct field-theoretic derivation of relation (\ref{cv1}) in 3+1D. Despite the correspondence apparent within geometric 
engineering, discussed in \S2,
the general proof of the Cecotti-Vafa relation in 1+1D does not at first sight appear to have a natural physical rationale in 3+1D.  It would be particularly 
interesting were a 
purely four-dimensional field-theoretic perspective to be placed on the Cecotti-Vafa relation, not least because it may be that 
many other aspects of the ensuing classification program \cite{cv} can be translated to \ntwo gauge theories in a similar way.

\subsection*{Acknowledgements}
I'd like to thank M.~Shifman, D.~Tong, and A.~Vainshtein for very helpful discussions and comments on the manuscript.
This work was supported in part by NSERC, Canada.

\end{document}